\documentclass[traditabstract]{aa} 
\usepackage{epsfig}
\usepackage{graphicx}
\usepackage{xcolor}
\usepackage{txfonts}
\usepackage{amssymb}
\usepackage{natbib}

\bibpunct{(}{)}{;}{a}{}{,}
\newcommand{\be}{\begin{equation}}
\newcommand{\ee}{\end{equation}}
\newcommand{\bea}{\begin{eqnarray}}
\newcommand{\eea}{\end{eqnarray}}

\usepackage{ulem}

\begin{document}

\title{Comparison of the disk precession models with the photometric behavior of TT Ari in 2021-2023}
\author{
V.~F. Suleimanov\inst{1}
\and
K.~V. Belyakov\inst{2,3}
\and
J.~M. Perales\inst{3,4}
\and
V.~V. Neustroev\inst{5}
}


\institute{
Institut f\"ur Astronomie und Astrophysik, Kepler Center for Astro and
Particle Physics, Universit\"at T\"ubingen, Sand 1, 72076 T\"ubingen, Germany\\ \email{suleimanov@astro.uni-tuebingen.de}
\and Parallax Enterprise, Kazan, Russia
\and American Association of Variable Star Observers, 49 Bay State Road, Cambridge, MA 02138, USA
\and AstroMallorca, Palma, Balearic Islands, Spain
\and Space Physics and Astronomy research unit, PO Box 3000, FIN-90014 University of Oulu, Finland
}

\date{Received xxx / Accepted xxx}

   \authorrunning{Suleimanov et al.}
   \titlerunning{Comparison of the disk precession models  
    with the photometric behavior of TT Ari.}

\abstract
{We present a comparative analysis of photometric observations of the cataclysmic variable TT Ari in its bright state, obtained by the TESS orbital observatory in 2021 and 2023 and by ground-based amateur telescopes in 2022. The light curves from 2021 and 2022 are dominated by modulations with a period slightly shorter than the orbital one (negative superhumps), 0.13292 and 0.13273\,d respectively. In 2023, much stronger modulations appeared on a much longer time scale of a few days with an amplitude of up to 0.5 mag, compared to 0.2 mag in 2021. The negative superhump variability with the period of 0.1338\,d was also found  in the 2023 observations, but the significance of these negative superhumps is much lower than in the previous seasons. Less significant additional modulations with a period exceeding the orbital one (positive superhumps) were detected in 2021 and 2022. Their periods were 0.15106 and 0.1523~d, respectively. We also found a previously unnoticed periodic signal corresponding to the orbital period of 0.13755~d in the TESS observations in 2021. Theoretical models of tidal precession of an elliptical disk predict a decrease in the precession period (and an increase in the positive superhumps period) with increasing disk radius, which is consistent with the observed photometric behavior of the system. It enables us to estimate the mass ratio of the components in TT~Ari to be $q$ in the range 0.24-0.29. The tilted disk precession model predicts a period of nodal precession whose value is in general agreement with observations.}


\maketitle
%

\section{Introduction}

TT Ari is a bright (about 10.5 - 11\,mag in the V band) cataclysmic variable star belonging to the subclass of the VY Scl stars (anti-dwarf novae), and is available for systematic observations even with small amateur telescopes. This is a binary system consisting of a white dwarf and a late-type  (M3.5) main sequence star \citep {1999A&A...347..178G}. The orbital period $P_{\rm orb}$ of TT Ari, determined from the radial velocity curves, is 0.13755 days \citep {1975ApJ...195..413C}. Most of the time the system is in a bright (high) state. Sometimes the brightness of the system decreases  from 11 to 17\,mag, as in 1983 and 2009 \citep {2010SASS...29...83K}.

Estimates of the mass ratio of the components $q$ and the inclination angle of the orbital plane to the line of sight $i$ are within the limits:
0.19  $ \le q=M_2/M_1 \le$  0.4; 17$^\circ \le$ i $\le$ 30$^\circ$ according to \citet{1975ApJ...195..413C}, \citet{2002ApJ...569..418W}, and \citet{ 2010AIPCB}. Here $M_2$ is the mass of the secondary donor star, and $M_1$ is the mass of the white dwarf. The system is not eclipsing because of the small inclination angle. However, the brightness of the system varies with an amplitude of a few tenths of a magnitude with a period close to the orbital, but noticeably different from it. This brightness variability was discovered in 1969 \citep{1969CoKon..65..355S} and studied further by other authors
\citep[see, for example][]{1987AcAS,1999AJA,2013AcA....63..453S, 2019MNRAS.489.2961B}.
It was shown that the photometric period can be either shorter or longer than the orbital one,
and it depends on the mean brightness of the system \citep{2001A&A...379..185S, 2013AstL...39..111B}.
If the brightness of the system lower than
V$\approx$ 11.3\,mag, then the photometric period exceeds the orbital one, while in the brighter state the photometric period is shorter than
orbital. Most of the time the system is in a bright state. In this aspect, the TT Ari system is similar to the novalike variable V603\,Aql, which demonstrates a photometric period longer than the orbital one, but when the system becomes brighter, it transits to a state with a photometric period shorter than the orbital \citep{1997PASP..109..468P, 2004AstLS}.

The reason for the photometric variability in such systems is still not reliably known. The most plausible is the model of an elliptical precessing accretion disk. In this case, the observed photometric period is the beat period between the orbital period and the disk precession period. Such a model was used to explain the photometric variability of SU UMa-type dwarf novae during especially bright and long outbursts \citep[so-called superhumps during superoutbursts,][]{1985A&A...144..369O}, and was confirmed by numerical calculations \citep{1988MNRASW}. According to these calculations, the disk becomes elliptical due to a 3:1 resonance between the Keplerian period at the outer edge of the disk and the orbital period. This kind of resonance is possible only for a sufficiently small mass ratio of the components $q\le 0.3$ \citep[or even $q<0.22$, see][]{2020AcA....70..313S}. An elliptical disk precesses in the direction of orbital motion (apsidal precession) and this kind of precession can explain only photometric periods exceeding the orbital one (positive superhumps).

\citet{1997PASP..109..468P} suggested that the accretion disk can be not only elliptical, but also inclined to the orbital plane. In this case, it is also precessing in the direction opposite to the orbital motion (nodal precession). In the restricted three-body problem  the period of the nodal precession  is about twice as long as the apsidal precession. If this relationship also exisits in the accretion disk precession, this should lead to a well-defined relationship between photometric periods longer and shorter than the orbital one if they are observed in the same system. 
Indeed, the ratio of the apsidal and nodal disk precession periods is close to 0.5 in the case of V603 Aql  \citep{1997PASP..109..468P, 2004AstLS}.

A search for photometric variability corresponding to the expected precession period of the
accretion disk has been  attempted by many authors \citep[see, for example,][]{1987AcAS, 1988AcAU, 1999A&AK}. The most complete analysis of the system's photometric behavior, summarizing its observations made over the past 40 years, is presented by \citet{2019MNRAS.489.2961B}. He showed, in particular, that the system's light curves often contain a periodicity that is close to the expected disk precession period, but does not coincide with it. \citet{2022MNRAS.514.4718B} has also performed an analysis of the long (quasi) continuous photometric observations of the system obtained by the TESS orbital observatory \citep{2014SPIE.9143E..20R}. He showed that in addition to the negative superhumps (predominant ones) and positive superhumps, only the beat period between these two photometric periods is reliably detected.

In addition to near-orbital photometric variability, TT\,Ari exhibits quasi-periodic oscillations with an amplitude about 0.1\,mag and a period close to 20 minutes \citep{1987AcAS, 1988AcAU, 1996A&AT, 1999A&AK, 2019MNRAS.489.2961B}. The nature of this kind of oscillations has not yet been established. At the same time, other authors do not find significant quasi-periodic oscillations in the range of 15-20 minutes \citep{2013AN....334.1101V} and note that such oscillations are transient short-lived phenomena \citep{2014AcAS}. There is no doubt, however,  that there is an excess in the power spectrum of the light curves against a background of red noise in the period range of 10-60 minutes \citep{1999A&AK, 2013AstL...39..111B}.

The present work aims to analyze two long quasi-continuous observations of the cataclysmic variable TT Ari obtained by the TESS observatory in 2021 and 2023, supplemented by our ground-based observations performed in 2022. We search for periodicities associated with the precession of the accretion disk and study the brightness variability of the system at time scales of tens of minutes.

\section{Observations}

The Transiting Exoplanet Survey Satellite (TESS) orbital observatory is designed primarily for the exoplanet searching.
However, as a result of its work, relatively long homogeneous time-series of observations of a large number of stars,
including cataclysmic variables, become available. In particular, TESS observed TT~Ari twice: once in 2021 from August 20
to October 10 in Sectors 42 and 43, and once very recently in 2023 from September 20 to November 11 in Sectors 70 and 71.
An analysis of the first TESS observation was presented in \citet{2022MNRAS.514.4718B}. Here we re-examine this TESS light
curve and analyse the second light curve. We then compare the findings with the analysis of our own observations.
Throughout the paper the TESS data sets are referred to as TESS-21 and TESS-23. In the following section we show that
during the TESS-23 observations, around JD 2460240, the photometric behaviour of TT~Aur has changed dramatically. For this
reason, we divided the TESS-23 set into two subsets, TESS-23a and TESS-23b. Also, after a short gap in the TESS-21 light curve between JD 2459473 and 2459474, the average brightness of TT~Ari decreased by about 0.05 mag, which has been accompanied by a somewhat change in the character of the variability (see below). These first- and second-half pieces of the TESS-21 light curve are referred to as TESS-21a and TESS-21b.
For the convenience of comparison of the TESS data with our own observations, we converted the TESS fluxes $F$, given in units of
electrons per second, into TESS magnitudes $T$ using the relation $T = -2.5\log_{10} F + ZP$ where $ZP$ = 20.44 is the TESS
Zero Point magnitude based on the values quoted in the TESS Instrument Handbook.

\begin{figure}
\centering
\includegraphics[width=0.9\columnwidth]{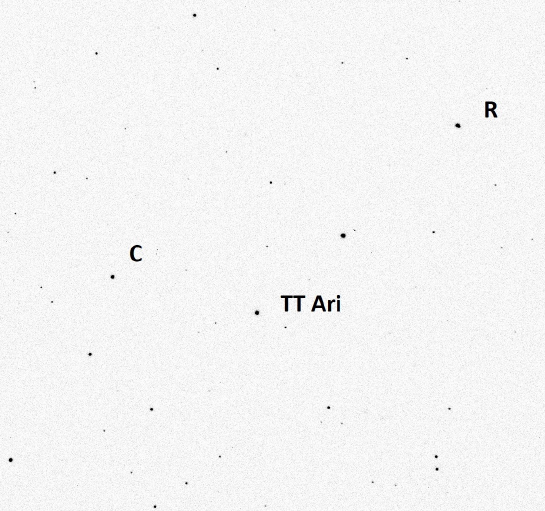}
\caption{A finding chart of TT Ari.}
  \label{fig1}
\end{figure}

\begin{table}
\caption {Comparison stars} \label{tab1}
\medskip
\begin{tabular}{p{0.4in}|p{1.15in}|p{1.15in}}

\hline \hline
Star & R:~~TYC 1207-1535-1  & C:~~TYC 1207-1413-1
\\
\hline
$\alpha $${}_{2000}$ & 02 06 18 &  02 07 14  \\
$\delta $${}_{2000}$ & +15 23 22 &  +15 19 50 \\
V  & 11.024 & 11.829 \\
\hline \hline
\end{tabular}\\
\end{table}

\begin{figure*}
\includegraphics[width=1.91\columnwidth]{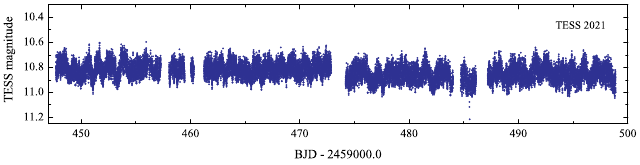}
\includegraphics[width=1.9\columnwidth]{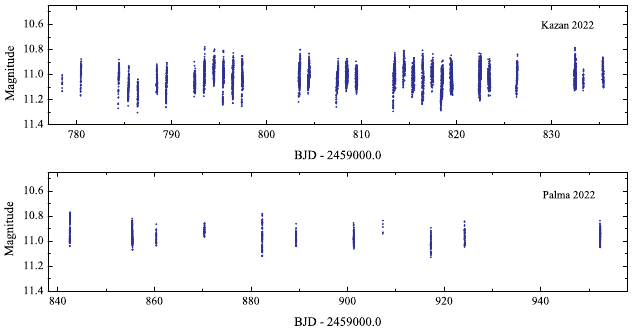}
\includegraphics[width=1.91\columnwidth]{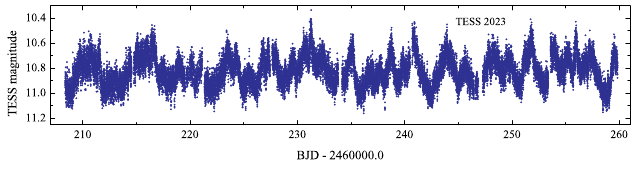}
\caption{Light curves of TT Ari obtained by TESS in 2021 and 2023 (top and bottom panels, respectively), 
and in Kazan and Palma de Mallorca in 2022 (two middle panels).
}
  \label{fig2}
\end{figure*}

Additional observations of TT Ari were taken with two amateur telescopes from July 2022 to January 2023;
a total of 44 nights of data with a total duration of 136 hours were obtained.
Observations from 2022 July 17 to 2022 September 16 were carried out using a 5" Schmidt-Cassegrain telescope, installed 15 km
northeast of Kazan (Russia) in an area free from urban development and ambient light sources. The CCD camera ZWO ASI-294MC
without filters was used as a detector. The duration of each observation was 20-40 seconds. Primary reduction was performed
by the standard method using MaxIm DL 5.10 software.
The following observations from 2022 September 20 to 2023 January 07 were performed in Palma (Majorca, Balearic Islands, Spain)
in a zone of urban development and strong light pollution. A 127-mm Maksutov-Cassegrain telescope with the $V$ filter
was used with the CCD camera ZWO ASI533MM. The exposure time of each observation was 60 seconds. Primary reduction was performed
using DeepSkyStacker 3.3.2.

Aperture photometry was carried out using the nearby comparison (R) and check (C) stars whose brightness was taken from the AAVSO database.
These stars are indicated in a finding chart of TT~Ari in Fig.\ref{fig1}, and their $V$ magnitudes are listed in Table~\ref{tab1}.

\section{Analysis of the variability of TT Ari }

Light curves of TT Ari are shown in Fig.\,\ref{fig2}.  In 2021-2023, when the analyzed data 
were collected, the system was in a bright state varying between roughly 10.6 and  11.2\,mag. We note that the variability of the system in 2023 differs significantly from its usual photometric behavior. Usually, the variability of TT Ari dominates by a photometric period close to the orbital one \citep[see, e.g.][]{2019MNRAS.489.2961B}. However, the dominant variability time scale was significantly longer in 2023, about a few days.  

Using the Generalised Lomb-Scargle (GLS) period analysis method \citep{2009A&A...496..577Z} as implemented in
PyAstronomy\footnote{https://github.com/sczesla/PyAstronomy} \citep{pyAst}, we calculated the power spectra of individual data sets.
These periodograms exhibit several strong peaks which can be roughly divided into two groups: located around the orbital frequency
$\nu_{\rm orb}$=7.27~d$^{-1}$, and below than $\sim$1~d$^{-1}$.
These two groups of periods are investigated further separately.
The period errors, given below, were also estimated by the PyAstronomy/GLS method as it delivers an associated error when estimating the period.

To examine the evolution of signals over short time intervals,
we also calculated slidograms (i.e. sets of the power spectra calculated with the window moving along the time
series) of selected data sets, following the technique of a sliding periodogram \citep[see, e.g.][]{2005MNRAS.362.1472N}.
The slidograms were calculated adopting different values of the time-span of the window, which sets the frequency resolution
of the power spectrum. For frequencies close to and higher than the orbital one, we settled on the windows of 2-4 days, while
for low frequencies we used the window of 10 days as a good compromise between time resolution and frequency resolution.
More homogeneous nature of the
TESS data resulted in a higher quality periodograms which we analyze first and then compare the results with the two other data sets.

\begin{figure}
\includegraphics[width=0.49\columnwidth]{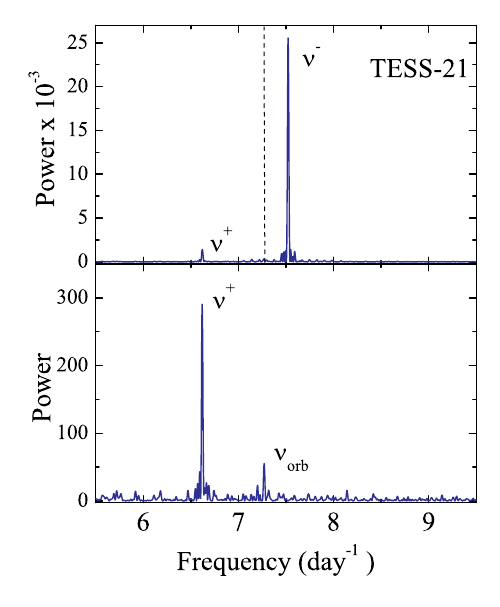}
\includegraphics[width=0.51\columnwidth]{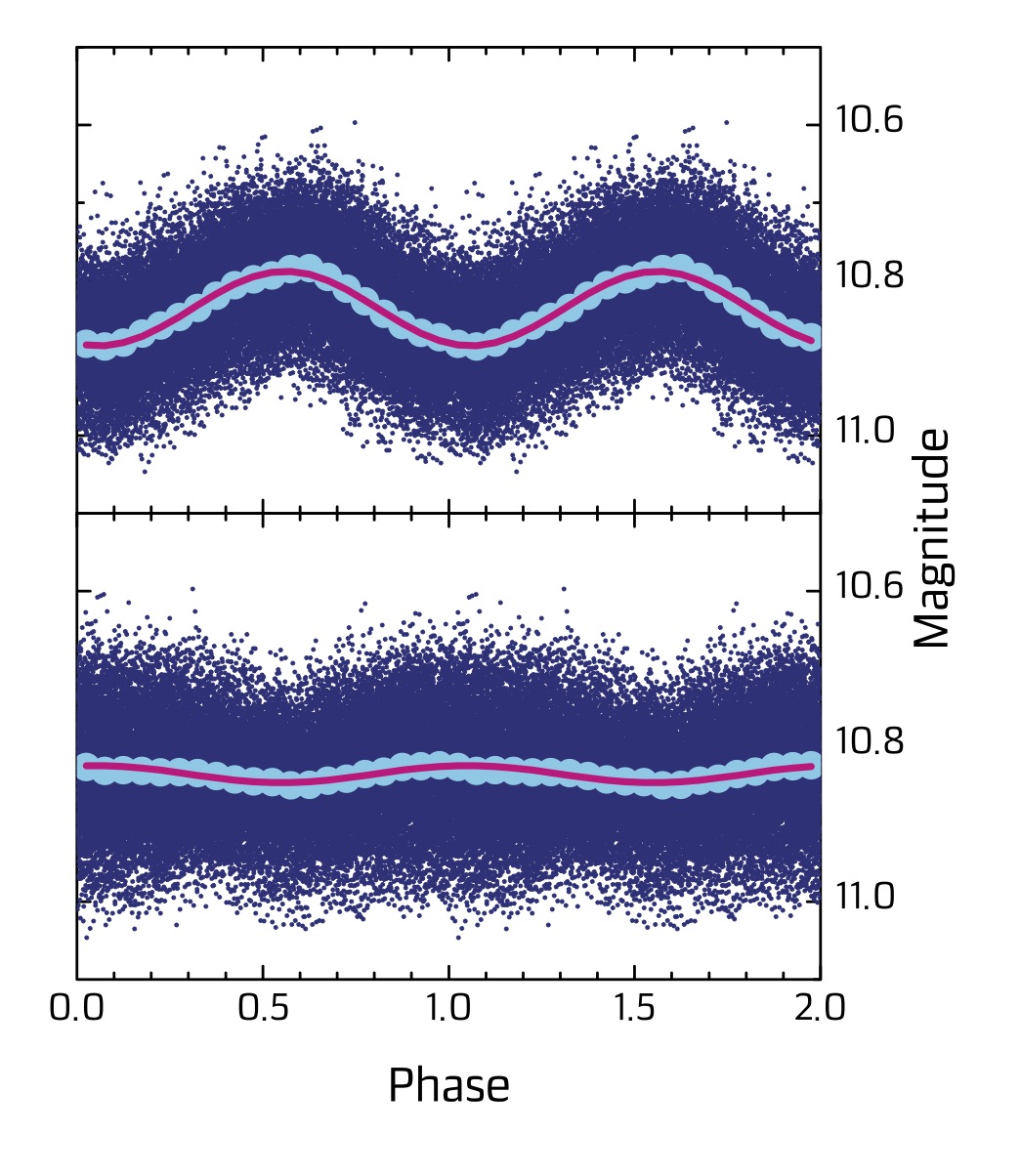}
\caption{\textit{Left panels:} Periodograms of the observed (top panel) and pre-whitened by the $\nu^{-}$ frequency (bottom panel) light curves of the TESS-21  observations in the frequency range of 5-10\,d$^{-1}$. The strongest peaks corresponding to the negative ($\nu^{-}$) and positive ($\nu^{+}$) superhumps and the orbital variability ($\nu_{\rm orb}$) are labeled. The position of the $\nu_{\rm orb}$ frequency in the top panel is shown by the vertical dashed line.
\textit{Right panels:} The TESS-21 light curve folded with the negative (top panel) and positive (bottom panel) superhumps periods. The phase-binned light curves and their best sinusoid fits are also shown.}
  \label{fig3}
\end{figure}

\begin{figure}
\includegraphics[width=0.99\columnwidth]{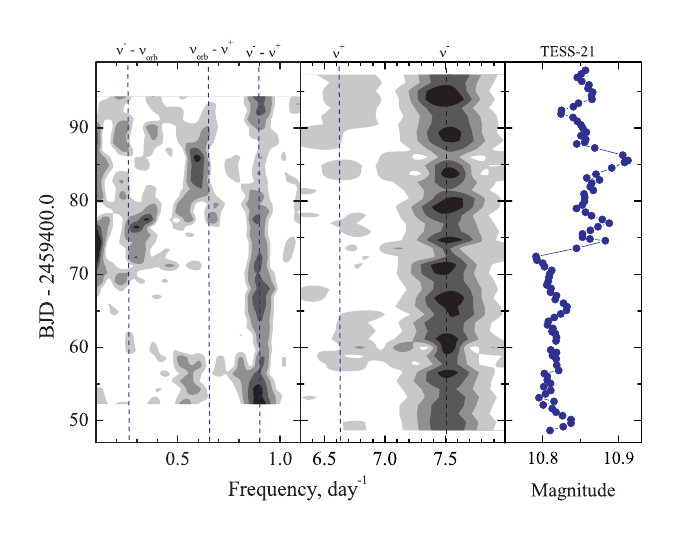}
\hspace{3mm}
\caption{
The sliding periodograms (slidograms) in two frequency ranges calculated for the TESS-21 data set, shown as grey-scale images. The slidograms in a low-frequency range of 0.1--1.1 d$^{-1}$ are calculated with a window of 10 days, while for frequencies close to the orbital one, we used the window of 2 days. The right panel show the light curve, also smoothed with the window of 2 days.
Dashed vertical lines in the slidograms denote the important frequencies discussed in the paper (see text for explanation).} 
 \label{fig4}
\end{figure}

The calculated slidograms were also used to investigate how amplitudes of selected variabilities depend on time during the TESS-21 and TESS-23 observations. To find these amplitudes, the light curves of all the time intervals (windows) were fitted by a sinusoid with the given period.

\subsection{Photometric periods close to the orbital period }

\citet{2022MNRAS.514.4718B} has noticed that the remarkable feature of the TESS-21 light curve is the simultaneous presence of negative
and positive superhumps (modulations with periods shorter and longer than the orbital one, respectively), something that has never
previously been observed in TT~Ari. We confirm (see Fig.\,\ref{fig3}, top left panel) that TESS-21 is indeed dominated by negative superhumps with the period of 0.132920(2)~d
($\nu^{-}$=7.5233(1)~d$^{-1}$ and at least three first harmonics) and the total amplitude of 0.05 mag
(see Fig.\,\ref{fig3}, top right panel, and Table\,\ref{tab:per}).
 Positive superhumps with the period of 0.15106(1)~d are evident in
the power spectrum as the peaks at $\nu^{+}$=6.6200(6)~d$^{-1}$ (see Fig.\,\ref{fig3}, bottom left panel) and its first harmonic 2$\nu^{+}$.  
The amplitude of this periodicity is significantly smaller, about 0.01 mag (Fig.\,\ref{fig3}, bottom right panel). The beat between these two signals
($\nu^{-}$-$\nu^{+}$) produces a very strong peak at 0.9011(3)~d$^{-1}$ (1.1094(3)~d) and peaks at its first two harmonics. 
This period is the second most important with an oscillation amplitude of approximately 0.03 mag, see more detail consideration in the next subsection.
There is also evidence for a signal at $\nu^{-}$+$\nu^{+}$ = 14.143~d$^{-1}$. Reliable detection of two photometric periods, above and below
the orbital one, indicates that both the apsidal and the nodal precessions were found simultaneously. It means that during the observational
interval the disk was elliptical, and at the same time inclined to the orbital plane, if the model suggested by \citet{1997PASP..109..468P} is correct.

\begin{figure}
\includegraphics[width=0.99\columnwidth]{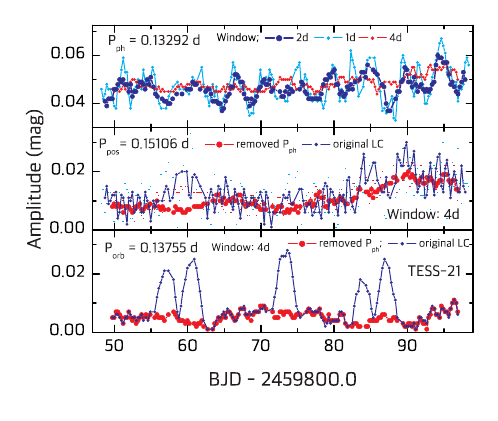}
\includegraphics[width=0.99\columnwidth]{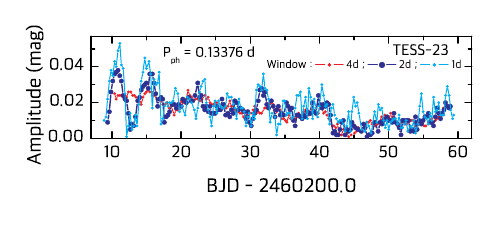}
\caption{Time dependencies of amplitudes of found photometric modulations with frequencies higher than 1\,d$^{-1}$ for TESS-21 (top panels) and TESS-23 (bottom panel) data sets.}
  \label{fig6a}
\end{figure}

Surprisingly, we also detect a peak
at $\nu_{\rm orb}$=7.271(1)~d$^{-1}$ (0.13753(3)~d) corresponding to modulations with the orbital period. Although this peak is relatively
weak in the original periodogram, it becomes apparent after pre-whitening the light curve with the negative superhumps' signal (Fig.\,\ref{fig3},  bottom left panel). The amplitude of the variability with this period is very low, about 0.005 mag. In the past,
orbital photometric modulations were detected only during the deep low state and have never been seen in the high state
\citep{2019MNRAS.489.2961B,2022MNRAS.514.4718B}.

\begin{figure}
\includegraphics[width=0.49\columnwidth]{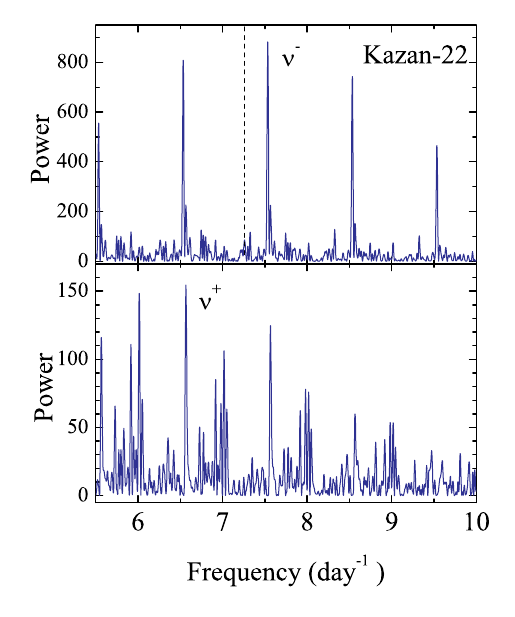}
\includegraphics[width=0.50\columnwidth]{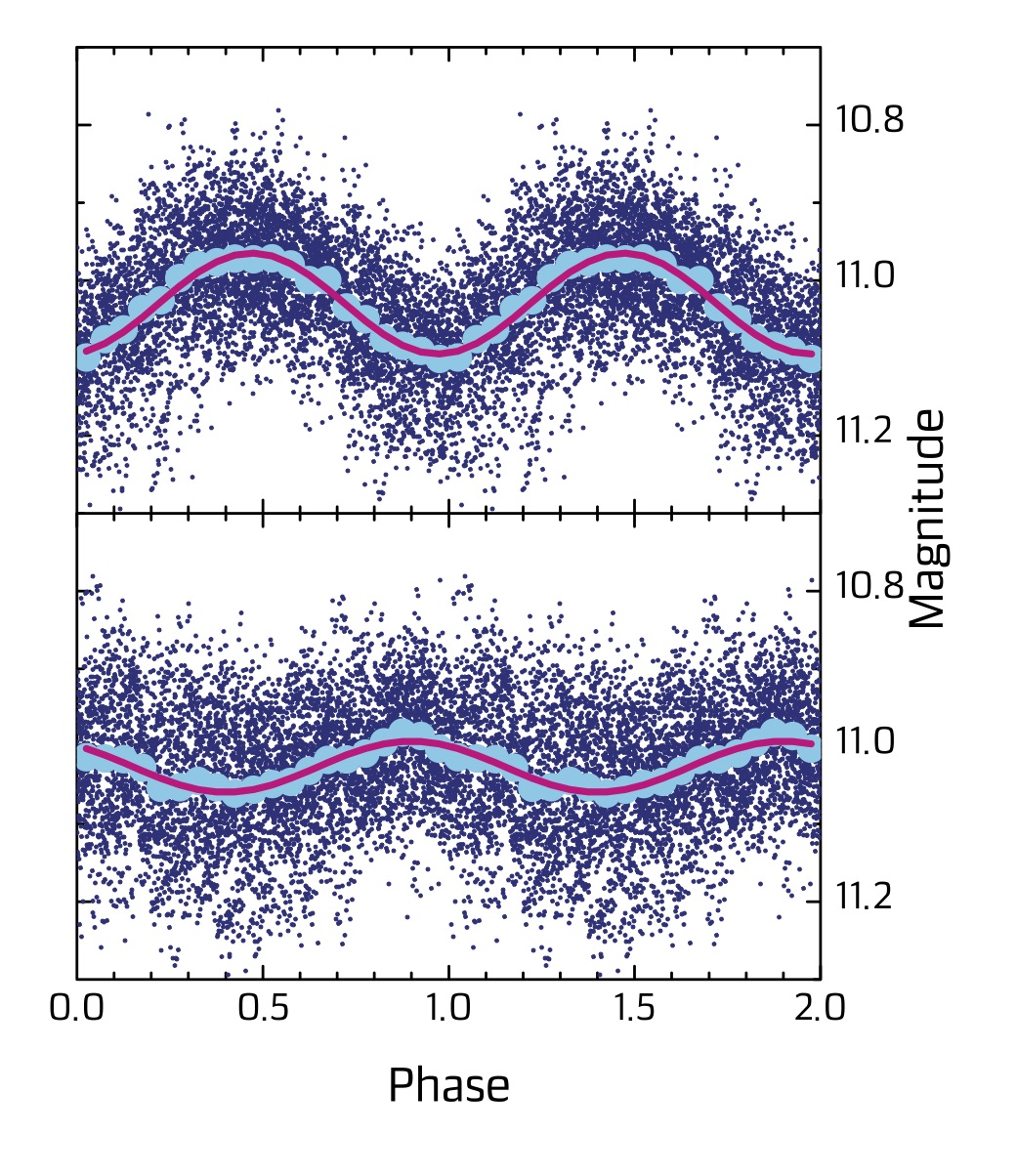}
\caption{
The same as in Fig\,\ref{fig3}, but for the Kazan observations.}
  \label{fig3b}
\end{figure}

\begin{figure}
\includegraphics[width=0.49\columnwidth]{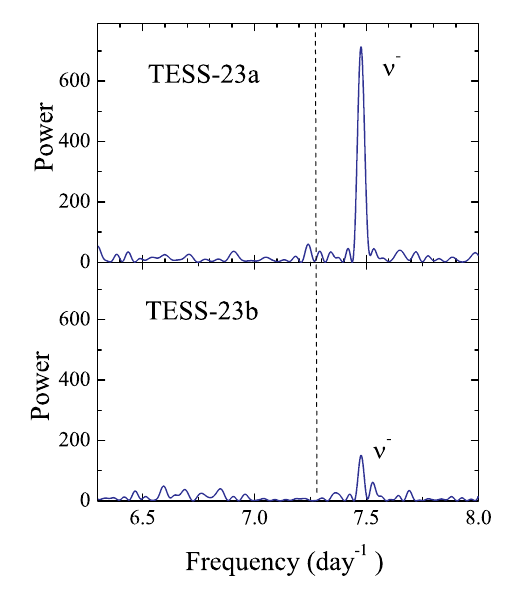}
\includegraphics[width=0.50\columnwidth]{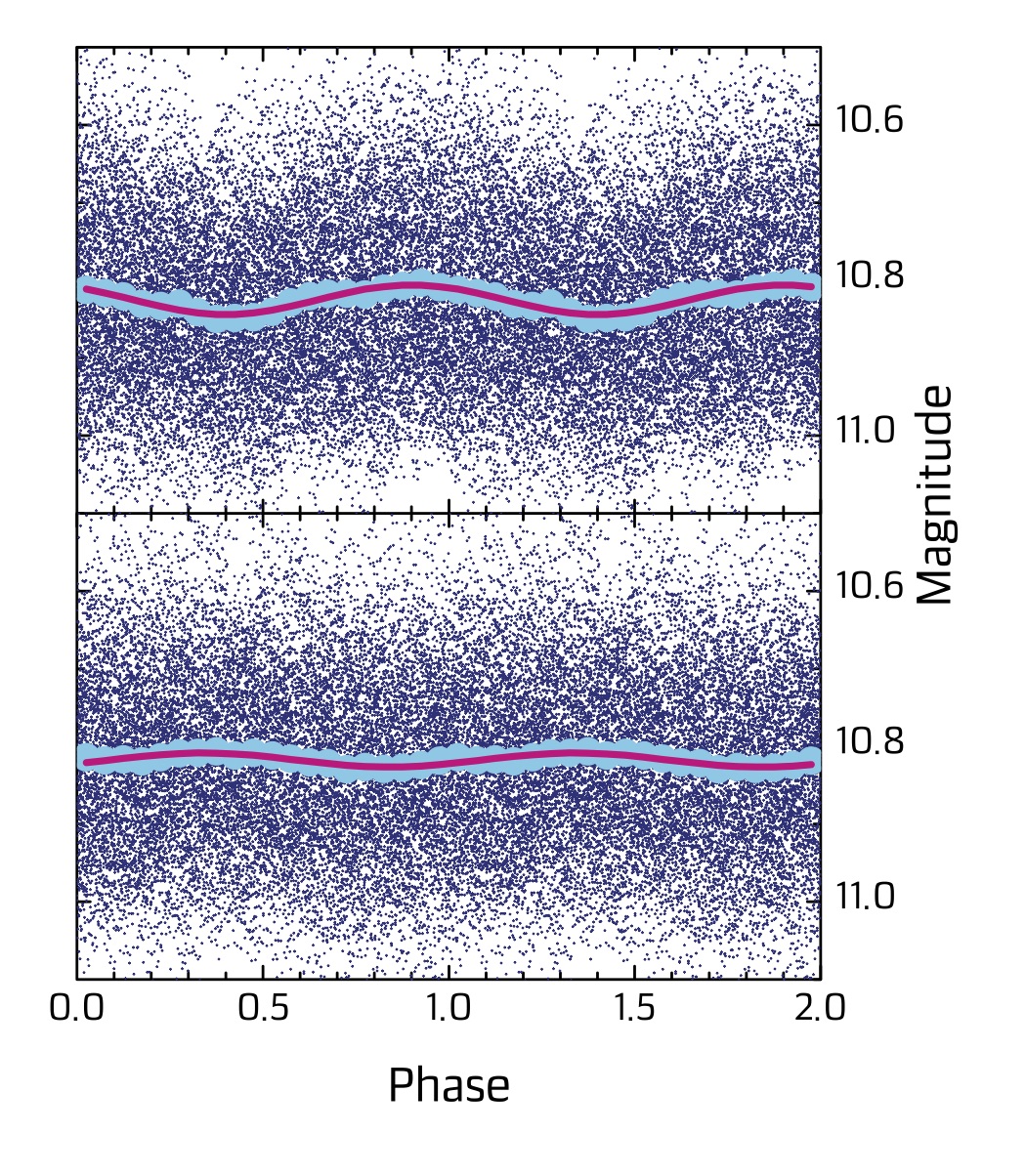}
\caption{\textit{Left panels:}
Periodograms of the observed light curves of the TESS-23a (top panel) and TESS-23b (bottom panel) observations in the frequency range of 6-8\,d$^{-1}$. The strongest peak corresponding to the negative ($\nu^{-}$) superhumps is labeled. The position of the $\nu_{\rm orb}$ frequency  is shown by the vertical dashed line.
\textit{Right panels:} The TESS-23a (top panel) and TESS-23b (bottom panel) light curves folded with the negative superhumps' period. The phase-binned light curves and
their best sinusoid fits are also shown.}
  \label{fig3c}
\end{figure}

It is obvious that the $\nu^{+}$ peak is much less significant than that of $\nu^{-}$. Moreover, the slidogram of TESS-21 (Fig.\,\ref{fig4}) shows that the positive superhumps were not always visible, and were most significant during the last 10 days of this observation. 
We also note that the significance of the peak corresponding to the negative superhumps varies during the observation interval of about four days, which is close to the expected period of the inclined accretion disk precession. 

An analysis of the time dependencies of superhump amplitudes during the TESS-21 observation confirms these conclusions 
(Fig.\,\ref{fig6a}, top panels). Indeed, when using a sliding window of 1 or 2 days, the amplitude of the negative superhumps shows variability with a typical time scale of 4-5 days. These variations are smoothed at the longer windows.  The amplitudes of the positive superhumps increase at the end of the observational set and show variability with the period corresponding to the beat frequency with the negative superhumps period $\nu^{-}$-$\nu^{+}$. These beats disappear if the $\nu^{-}$ pre-whitened light curve is considered.
Similar irregular resonances with the negative superhumps are seen in the time dependence of the orbital variability amplitude. They also disappear if the pre-whitened light curve is considered.

The periodogram of the Kazan observations obtained roughly 300 days after TESS-21 is generally consistent with that of TESS-21,
despite being affected by aliases due to gaps and irregularities present in the data sampling (see Fig.\ref{fig3b}, top left panel). This power spectrum is also dominated by the
very strong peak at $\nu^{-}$=7.5340(3)~d$^{-1}$ and its 1-day aliases (2$\nu^{-}$ is also present). After pre-whitening the light curve with
this signal, the positive superhumps' signal at $\nu^{+}$=6.5628(9)~d$^{-1}$ becomes apparent 
(Fig.\ref{fig3b}, bottom left panel).
The amplitudes of these periodicities are larger than in 2021, 0.065 mag and 0.033 mag, respectively (Fig.\ref{fig3b}, right panels). However, the peak corresponded to the beat between these two signals $\nu^{-}$-$\nu^{+}$=0.971~d$^{-1}$ is very weak, in contrast to TESS-21, where it appeared to be the second strongest in the periodogram.

Unfortunately, the following observations made in Palma have a large duty cycle and relatively short durations during the nights. Therefore,
the corresponding periodogram is heavily affected by an extremely complex window function with large aliasing effects, making it difficult
to reliably conclude the exact frequencies of dominated oscillations. However, the frequency of the strongest peak at 7.5273(3)~d$^{-1}$ is very close to $\nu^{-}$ from the Kazan data.

\begin{figure}
\includegraphics[width=0.99\columnwidth]{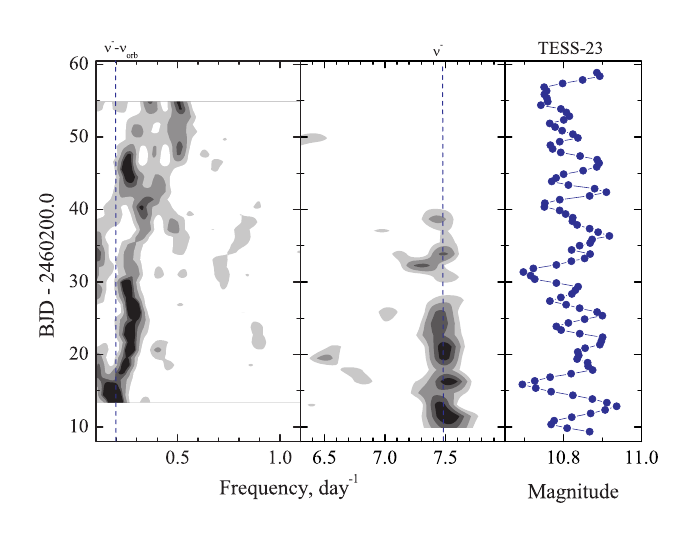}
\caption{
The same as Fig.\,\ref{fig4}, but for the TESS-23 observations.} 
 \label{fig4a}
\end{figure}

Interestingly, the appearance of the light curve of the TESS-23 data set and its corresponding periodogram is completely different from TESS-21.
The light curve demonstrates strong modulations (with an amplitude of up to 0.5 mag, compared to 0.2 mag in TESS-21) on a relatively long
(a few days) time-scale.
The power spectrum of the entire TESS-23 data set is indeed dominated by low-frequency modulations, which we discuss in Section~\ref{sec:prec},
and only shows a weak signal at $\nu^{-}$=7.476(18)~d$^{-1}$ (0.1338(3)) with a low amplitude of 0.01--0.02 mag (see
Fig.\,\ref{fig3c}).
The sliding periodogram in the frequency range close to the orbital one demonstrates (Fig.\,\ref{fig4a}) that on about JD 2460240 TT~Ari had experienced a transition to a state in which the negative superhumps, which were seen clearly before that date, suddenly almost disappeared.  As seen in Fig.\,\ref{fig6a} (bottom panel), the amplitude of the negative superhumps after this date dropped below 0.01 mag but then started increasing slowly again.

Due to the presence of such a transition, we split the TESS-23 set into two subsets, TESS-23a and TESS-23b, 
and studied them separately.  We find that their periodograms shown in Fig.~\ref{fig3c} are not notably different, apart from a much stronger $\nu^{-}$ signal in TESS-23a
compared to TESS-23b, and somewhat stronger low-frequency signals in TESS-23b than in TESS-23a (see Fig.\,\ref{fig7} below). 
However, even in TESS-23a the $\nu^{-}$ signal was much weaker than it was in our previous data sets.

\subsection{Low-frequency range: search for direct evidence of disk precession}
\label{sec:prec}

All the calculated periodograms show relatively strong peaks at frequencies $\lesssim$1~d$^{-1}$. In this interval, in addition to modulations
at the beat $\nu^{-}$-$\nu^{+}$ frequency, a direct periodic signal from the precession of the accretion disk is expected \citep[see, for
example][]{1987AcAS, 2013AcA....63..453S}. As mentioned before, photometric periods close to the orbital one, $P_{\rm ph}$, are interpreted
as beat periods between the orbital period $P_{\rm orb}$ and the disk precession period $P_{\rm pr}$
\be \label{eq:eq1}
  \frac{1}{P_{\rm ph}}=\frac{1}{P_{\rm orb}}\pm\frac{1}{P_{\rm pr}}.
\ee
  Here the minus sign is used when the disk precesses in the direction of the orbital motion (apsidal line precession, a positive superhump); the plus sign is used when the disk precesses in the opposite direction to the orbital motion (tilt disk node line precession, a negative superhump).

Using this expression, we can estimate the expected disk precession periods from the found photometric periods of superhumps.
It appeared that the period of precession of the line of apsides $P_{\rm pr}^+$  should have been close to 1.538 and 1.414~d
($\nu_{\rm pr}^+$=0.650 and 0.707~d$^{-1}$) at the time of the TESS-21 and Kazan observations, respectively, while the period
of precession of the line of nodes $P_{\rm pr}^-$ should have been 3.953, 3.788, and 4.831~d ($\nu_{\rm pr}^-$=0.253, 0.264, and 
0.207~d$^{-1}$) at the time of the TESS-21, Kazan, and TESS-23a observations, respectively.
We note that the derived ratio of precessions $P_{\rm pr}^+/P_{\rm pr}^-$ is almost constant, 0.39 and 0.37 for the
observations in 2021 and 2022. This ratio is smaller than was expected from the ratio of the precession periods for the Moon (0.47) and that found for V603\,Aql \citep[0.47-0.51,][]{1997PASP..109..468P, 2004AstLS}.

\begin{figure}
\includegraphics[width=0.49\columnwidth]{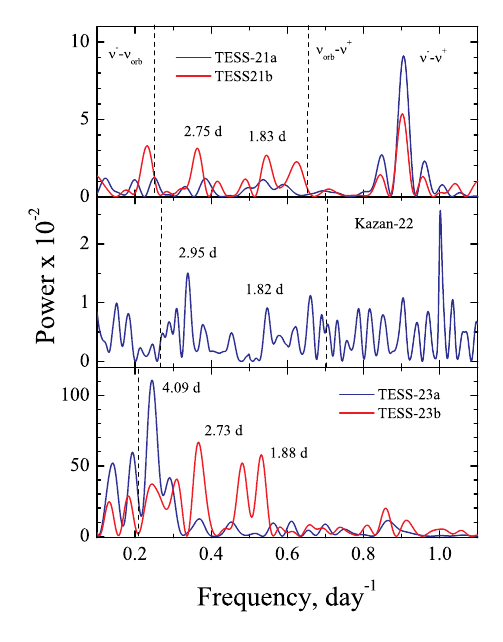}
\includegraphics[width=0.51\columnwidth]{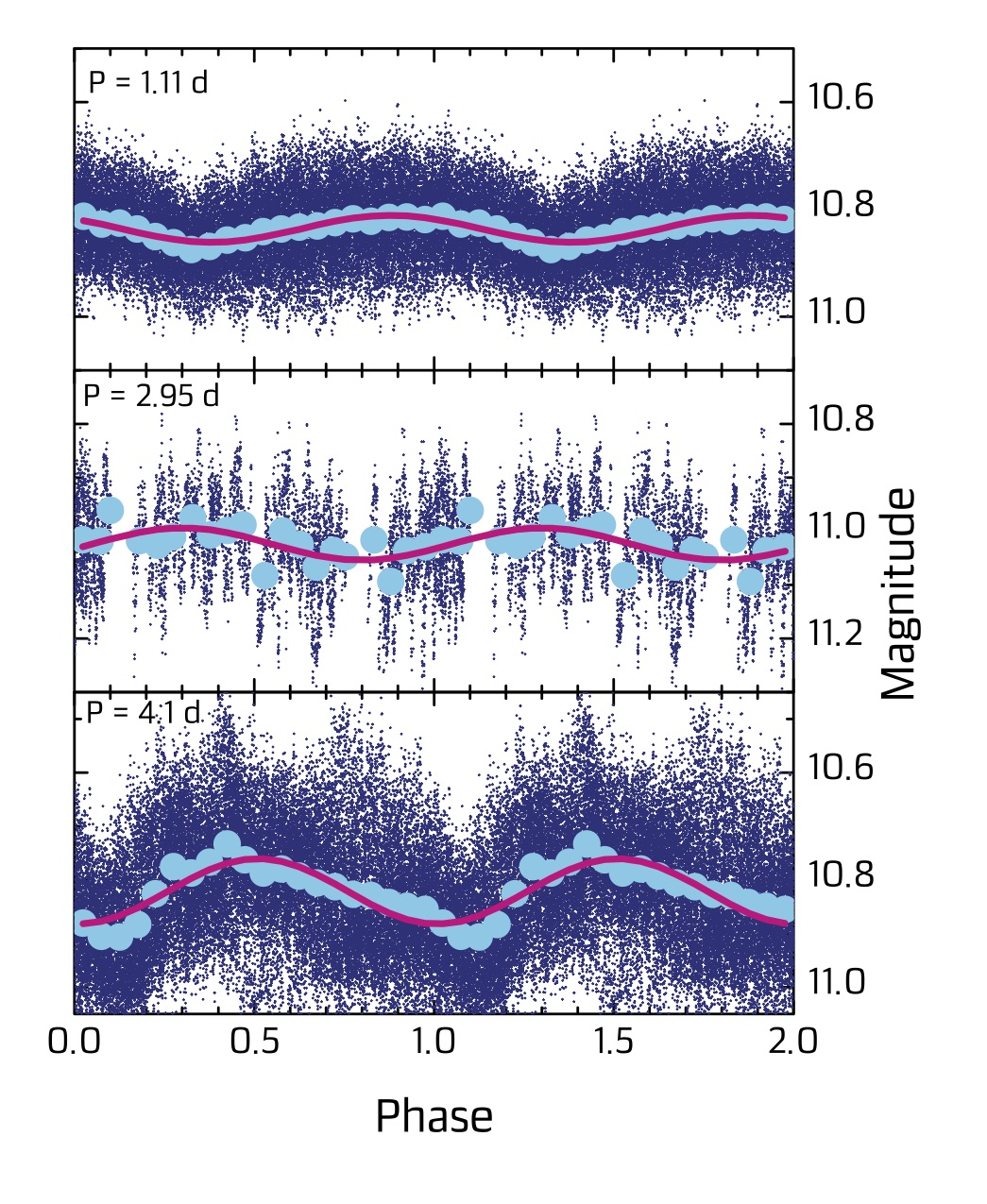}
\caption{\textit{Left panels:}
Periodograms of the light curves obtained during all the observational intervals in the frequency range of 0.1-1.1\,d$^{-1}$.
The vertical dashed lines show the positions of the frequencies corresponding to the expected periods of the nodal (left dashed lines) and apsidal (right dashed lines in the two upper panels) precessions of the tilted elliptical accretion disk. The periods corresponding to the two most stable peaks around periods 2.7-2.9\,d and 1.8-1.9\,d are marked. A peak corresponding to the observation duty cycle of 1 day is also visible in the middle panel. \textit{Right panels:} The light curves folded with the most significant low-frequency periods.}
  \label{fig7}
\end{figure}

Periodograms in a frequency range of 0.1-1.1\,d$^{-1}$ for all the investigated observational intervals are presented in 
Fig.\,\ref{fig7}, left panels. There are almost no clear peaks at the expected disk precession period, except for the peak 
at 0.23-0.25\,d$^{-1}$ in TESS-21 which corresponds to $\nu_{\rm pr}^-$ with the low amplitude about 0.01-0.02 mag. Although this peak is relatively weak, the sliding periodogram demonstrates (Fig.\,\ref{fig4}, left panel) that the $\nu_{\rm pr}^-$ signal appears much stronger after JD~2459474. Indeed, the peak close to  $\nu_{\rm pr}^-$ in the periodogram of TESS-21b 
is more significant (Fig.\,\ref{fig7}).

\begin{figure}
\includegraphics[width=0.99\columnwidth]{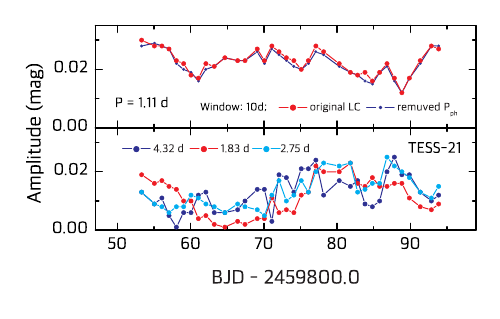}
\includegraphics[width=0.99\columnwidth]{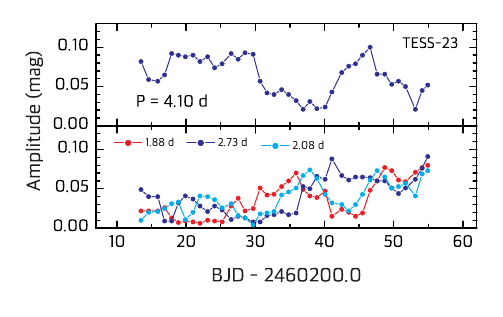}
\caption{
Time dependencies of amplitudes of found photometric modulations with frequencies lower than 1\,d$^{-1}$ for TESS-21 (top panels) and TESS-23 (bottom panel) data sets.
}
  \label{fig7a}
\end{figure}

It is interesting that the strengthening of the $\nu_{\rm pr}^-$ modulation in TESS-21b is accompanied by a weakening of the beat signal ($\nu^{-}$-$\nu^{+}$) at 0.901~d$^{-1}$, see also Fig.\,\ref{fig7a}, the top panels.
There was also an appearance of a weak signal with a frequency close to $\nu_{\rm pr}^+$=0.650~d$^{-1}$. 

In contrast to TESS-21, TESS-23 is dominated by low-frequency modulations which produce several strong peaks in the periodogram 
(Fig.\,\ref{fig7}, left panels). However, the corresponding slidogram (Fig.\,\ref{fig4a}, left panel) indicates that all these signals have a low degree of coherence. Nevertheless, during the first half of TESS-23 and also later in this time interval there was a very strong signal with a frequency close to $\nu_{\rm pr}^-$, namely 0.234\,d$^{-1}$.  
The amplitude of this periodicity is about 0.1 mag (see the right panels of Fig.\,\ref{fig7} and the bottom panels of  Fig.\,\ref{fig7a}). 

There are also a few other prominent peaks in the presented periodograms, whose frequencies are close to each other in different time intervals. They correspond to periods 2.75\,d and 1.81\,d (TESS-21), 2.95\,d and 1.82\,d (Kazan-22), and 2.73\,d and 
1.88\,d (TESS-23b), see Fig.\,\ref{fig7}. We note that amplitudes of these periodicities increase in the second halves of both the TESS observations, and that during the TESS-23 observation the amplitudes of low-frequency oscillations 
are a few times larger than during other observations (see Fig.\,\ref{fig7a}).  A period close to 2.95~d was also detected earlier by other observers \citep[e.g., $2.92$~d,][]{1999AJA}. Formally, a peak that corresponds to the beat frequency $\nu^{-}$-$\nu^{+} \approx 0.968$\,d$^{-1}$ (1.033\,d) exists in the periodogram of the Kazan observations (see Fig.\,\ref{fig7}), but the significance of this peak is low.
All the found periods in the investigated frequency range are collected in Table\,\ref{tab:per}.

\begin{table} \caption {Periods found in the observational intervals sequentially TESS-21a, TESS-21b, Kazan-22, TESS-23a and TESS-23b. The periods in different intervals are divided by double lines.
\label{tab:per}
}
{
\begin{center}
\begin{tabular}{l|l|l|l}
\hline \hline
Period, Day & Frequency, d$^{-1}$ & A$^a$ (mag)& Interpretation \\
\hline
   0.13756(5)   & 7.269(3) & 0.006 &$\approx \nu_{\rm orb}$ \\  
   0.132919(6)  & 7.5234(3) & 0.046&$\nu^-$ \\  
   0.15114(4)   & 6.616(2) & 0.009&$\nu^+$\\  
   1.1045(6)     & 0.9054(5) & 0.028&$\nu^- - \nu^+$ \\  
     3.99      & 0.250  &0.01 &$\approx \nu_{\rm pr}^-$ \\  
     \hline
     \hline 
   0.13755(6)   & 7.270(3) & 0.006 &$\approx \nu_{\rm orb}$ \\  
   0.132903(6)  & 7.5243(4) & 0.049&$\nu^-$ \\  
   0.15107(3)   & 6.620(1) & 0.012&$\nu^+$\\  
   1.1085(8)    & 0.9021(7) & 0.023&$\nu^- - \nu^+$ \\  
     4.32      & 0.2315  &0.017 &$\approx \nu_{\rm pr}^-$ \\  
     1.83     & 0.545 & 0.015&$\nu_1$ \\  
     2.75     & 0.364 &0.017  &$\nu_2$ \\
     \hline
     \hline 
   0.132740(6) & 7.5335(3) & 0.065&$\nu^-$ \\  
   0.15232(2) & 6.5651(10) &0.033 &$\nu^+$ \\  
     1.83 & 0.547 & 0.02&$\nu_{\rm 1}$ \\  
     2.95 & 0.339 &0.03  &$\nu_{\rm 2}$ \\  
     1.03 & 0.968 &0.017  &$\nu^- -\nu^+$ \\  
 \hline \hline 
   0.13374(1) & 7.4771(6) & 0.019&$\nu^-$ \\ 
     4.10      & 0.244  & 0.074&$\approx \nu_{\rm pr}^-$ \\  
\hline \hline 
   0.13380(6) & 7.4741(32) & 0.009 &$\nu^-$ \\       
   4.10      & 0.244  & 0.045&$\approx \nu_{\rm pr}^-$ \\  
     1.88 & 0.531 & 0.053&$\nu_{\rm 1}$ \\  
     2.73 & 0.366 & 0.062 &$\nu_{\rm 2}$ \\  
     2.08 & 0.482 & 0.051 &$\nu_{\rm 3}$ \\  
\hline
\end{tabular}
\end{center}
}
\textbf{Notes:} 
$\nu_{\rm orb}$ - orbital frequency and corresponding period found earlier spectroscopically \citep{1975ApJ...195..413C};
$\nu_{\rm pr}^-$ - frequency corresponding to the nodal precession period; $\nu^-$ is 
the frequency of the negative superhumps; $\nu^+$ is the frequency of the positive superhumps.
$a$ - amplitude of the sinusoid that best fits the folded light curve. 
\end{table}

\subsection{Quasi-periodic oscillations on a scale of tens of minutes}

\begin{figure}
\includegraphics[width=0.49\columnwidth]{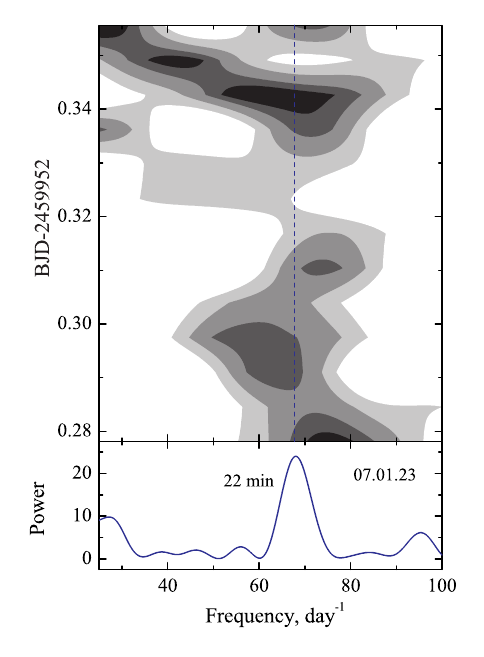}
\includegraphics[width=0.47\columnwidth]{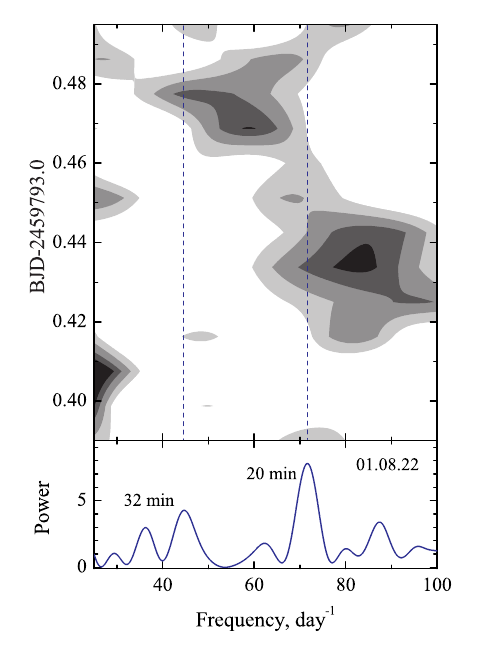}
\includegraphics[width=0.49\columnwidth]{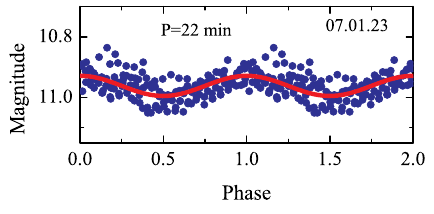}
\includegraphics[width=0.49\columnwidth]{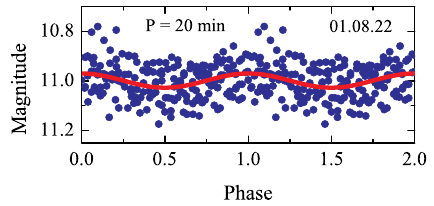}
\caption{\textit{Top panels}: Slidograms in the range of periods of 15-60 minutes during the nights 2023 January 7 (left panel) and 2022 August 1 (right panel).
\textit{Middle panels}: Corresponding periodograms in the same range of periods for the same nights.
\textit{Bottom panels:} Light curves folded with the most significant periods.}
  \label{fig8}
\end{figure}

A study of the quasi-periodicity of the system was fulfilled in the range of periods from 5 to 60 minutes using all the observed intervals. For the ground-based data (Kazan and Palma), such an analysis was conducted for each night with a duration of observation longer than 1.5 hours. The most frequently observed periods are in the range of 15$-$30 minutes with a brightness amplitude of about 0.02 mag. At the same time, we note that periods of about 15 and 27 minutes can be repeated over two or three consecutive nights. As an example,  Fig.\,\ref{fig8} (left-hand panel) shows the periodogram and how the periodogram changes during the night for the light curve obtained on 2023 January 7. It can be seen that the found period is not constant, but varies over times comparable with the period itself. At the end of the night, the frequency of this peak drifts towards lower frequencies. This is the most significant period found in the interval under study and it is close to the periods noted earlier as periods of quasi-periodic oscillations in the system \citep{1987AcAS}.

Another example, but for the night of 2022 August 1, is shown in the right panel of Fig.\,\ref{fig8}.
The periodogram of this night also shows a fairly significant period of 20 minutes. However, the corresponding slidogram demonstrates that no stable modulations with any characteristic period were present throughout the night. Moreover, the peaks in the night-averaged periodogram do not correspond exactly to the emerging short-period modulations seen in the slidogram.

\begin{figure}
\includegraphics[width=0.49\columnwidth]{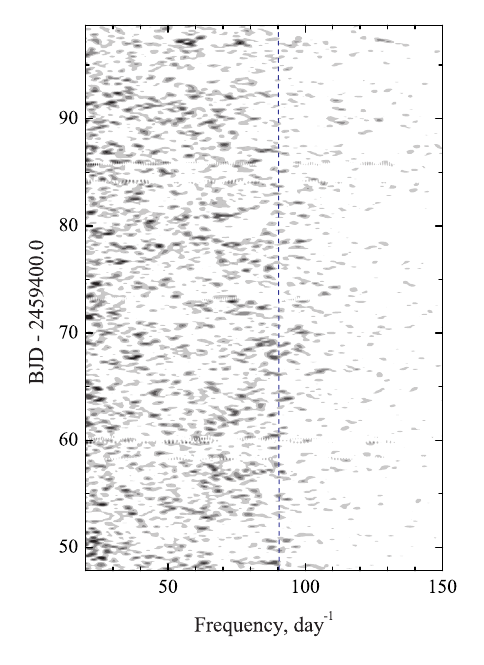}
\includegraphics[width=0.489\columnwidth]{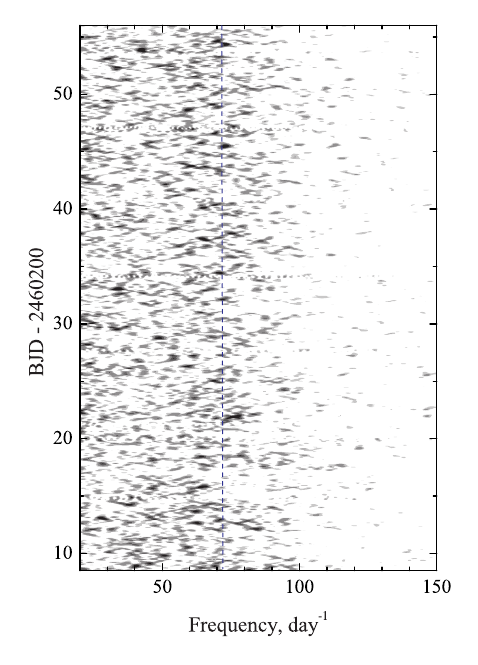}
\caption{Slidograms in the frequency range of 20-150\,d$^{-1}$ during the entire TESS-21 (left panel) and TESS-23 (right panel) observations. The vertical dashed lines show the positions of the break frequencies $\nu_c$,
90\,d$^{-1}$ (TESS-21) and 72\,d$^{-1}$ (TESS-23).
}
  \label{fig9}
\end{figure}

Results of the TESS observations are presented in the sliding periodograms in the range of periods of 15-60 minutes (Fig.\,\ref{fig9}). All the peaks appearing in this interval of periods are short-lived. 

\begin{figure}
\includegraphics[width=0.99\columnwidth]{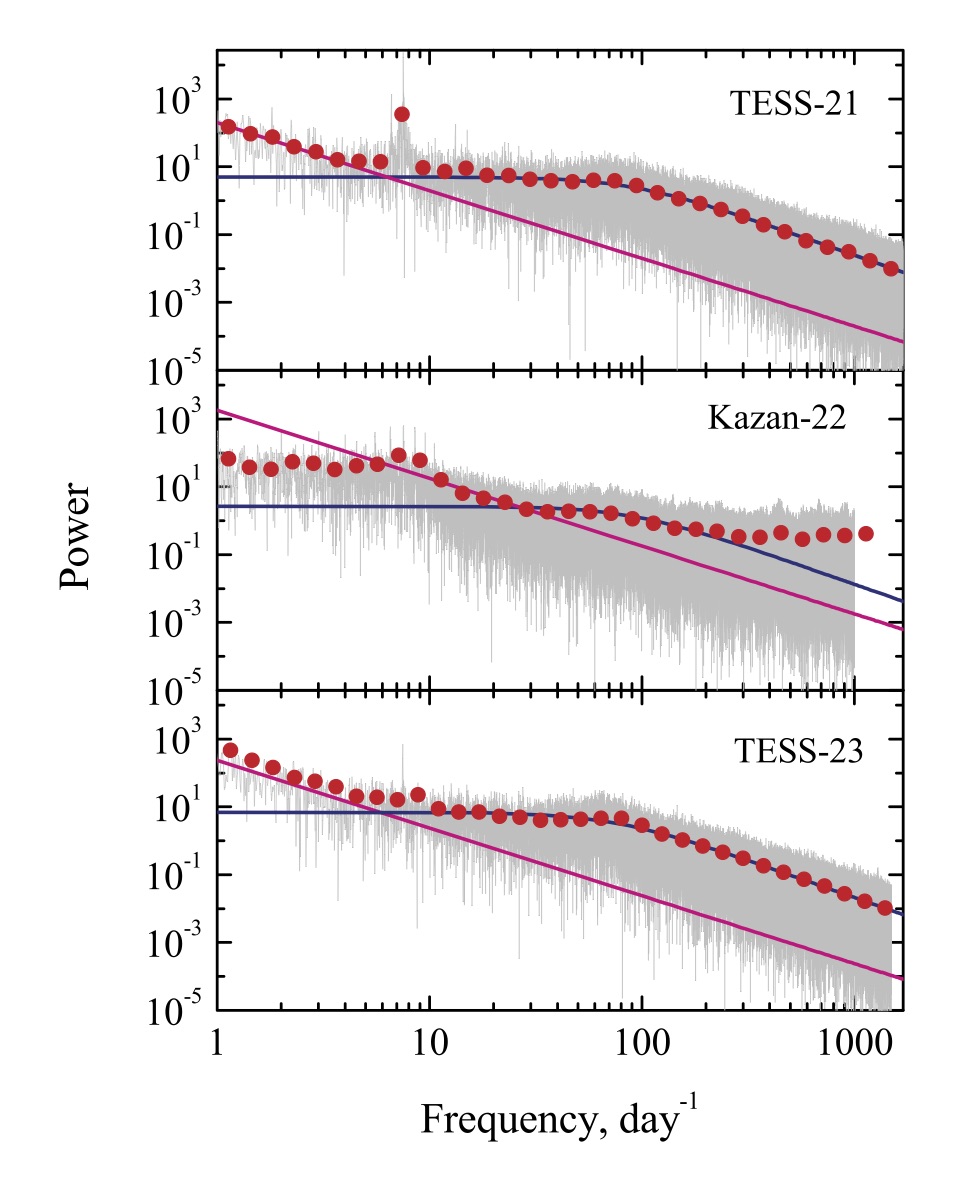}
\caption{Total power spectra (grey curves) in $\log$-$\log$ scale obtained using TESS-21 (top panel), Kazan-22 (middle panel), and TESS-23 (bottom panel) observations. The red circles show the averaged power spectra. The magenta lines shows the power law with an exponent of -2, and the blue curves are the fits of the excess power spectra, represented by the formula (\ref{eq:izb}) with the parameters $A=5$, $\nu_c = 90$\,d$^{ -1}$, $\gamma=2.2$ (TESS-21), $A=2.7$, $\nu_c = 90$\,d$^{ -1}$, $\gamma=2.2$ (Kazan-22), and $A=7$, $\nu_c = 72$\,d$^{ -1}$, $\gamma=2.2$ (TESS-23).
}
  \label{fig10}
\end{figure}

We also computed wide-range power spectra of the observed light curves corresponding to all the observed intervals, excluding Palma observations. The results are presented as separate power spectra in $\log$-$\log$ scale (Fig.\,\ref{fig10}).
Every power spectrum can be described by a power law (red noise), and
some additional noise component making the power spectrum flat at the frequencies between approximately 10\,d$^{-1}$ and  100\,d$^{-1}$.  The excess noise can be described by the analytical function
\be \label{eq:izb}
      P(\nu) = \frac{A}{1+(\nu/\nu_c)^\gamma}
\ee
\citep{1999A&AK}.

The red noise, which is probably arise due to random viscosity fluctuations in the accretion disk \citep{1997MNRAS.292..679L}, can  be approximated by the power law with an exponent of -2 in all the observational intervals. The exponent $\gamma=2.2$ describing the additional noise spectra, is also the same for
all the power spectra, but the power level in the TESS observations is higher, $A=5$ (TESS-21) and  $A=7$ (TESS-23) versus $A=2.7$ (Kazan-22). The break frequencies are identical for
power spectra obtained using observations in 2021 and 2022, $\nu_c$=90\,d$^{-1}$, but it appears lower in the TESS-23 observations, $\nu_c$=72\,d$^{-1}$.
 The higher accuracy of the TESS observatory data makes it possible to trace the red noise of the additional component up to the Nyquist frequency ($\approx 4000$ \,d$^{-1}$), and towards low frequencies down to the characteristic frequencies corresponding to several days.
The power spectrum of the Kazan observations is similar in appearance to the TESS power spectra. Due to limitations imposed by ground-based observations, the high frequencies above $\sim$200\,d$^{-1}$ are not available for analysis.

These results confirm the conclusion made earlier \citep{1999A&AK, 2013AcA....63..453S, 2019MNRAS.489.2961B},
that the system TT Ari shows no quasi-periodic brightness oscillations, and its variability at times of 10-60 minutes is stochastic noise, among which quite significant variability sometimes appears at some characteristic frequency from a given interval.

\section{Comparison with precession models and estimations of the system parameters} 

\subsection{Elliptical disk precession}
\label{s321}

According to the current concepts, the disk becomes elliptical and begins to precess towards orbital motion when it expands beyond the 3:1 resonance radius $R_{\rm 3:1}$ \citep{1988MNRASW, 1989PASJ...41.1005O}:
\be \label{eq:epsr1}
       \frac{R_{\rm 3:1}}{a} = r_{\rm res} = 3^{-2/3}(1+q)^{-1/3},
\ee
where $r_{\rm res}$ is a relative disk radius, and $a$ is the semi-major axis of the orbit \citep[formula (5.125) in][]{FKR02}. 

The period of precession $P_{\rm pr}^{+}$ and the corresponding period of positive superhumps $P_{\rm ph}^{+}$, to the formation of which the precession contributes, depend on the mass ratio $q$.
The superhump period excess $\varepsilon^+=(P_{\rm ph}^{+}-P_{\rm orb})/P_{\rm orb}$ can potentially be calibrated using accurate values of $q$ and observed values of $\varepsilon^+$. There were obtained several such empirical relationships between $\varepsilon^+$ and $q$, based on observations of cataclysmic variables (mostly of SU~UMa-type stars) showing positive superhumps \citep[see, e.g.,][and references therein]{2005PASP..117.1204P,2019MNRAS.486.5535M}. Using the dependence proposed by \citet{2005PASP..117.1204P}
\be \label{eq:epsq}
     \varepsilon^+ = 0.18\,q + 0.29\,q^2
\ee
and adopting the values of $\varepsilon^+$ observed in TT~Ari ($P_{\rm ph}^{+}$ varies between 0.1483 and 0.1523 days \citep[][and this work]{2019MNRAS.489.2961B}, therefore $\varepsilon^+$ is within 0.078-0.107), one obtains the mass ratio within 0.29 -- 0.37.

It must be noted that all the existing relationships are poorly calibrated at the larger $\varepsilon^+$ values end, which is populated by only a few systems with longer $P_{\rm orb}$ and higher $q$. These stars are not of SU~UMa-type but novalikes that display permanent superhumps, for which accurate measurements of system parameters, and in particular $q$, are not available. For example, four such systems from tables 3 and A1 in \citet{2019MNRAS.486.5535M} cover $\varepsilon^+$ in the range of 0.063--0.069 and measured to have $q$ between 0.24 and 0.31 with an accuracy of $\sim$20\%. Thus, to estimate $q$ in TT~Ari we have to rely on extrapolation. Despite this, the estimate of $q\approx$0.30 corresponding to $\varepsilon^+$=0.078 looks reasonable \citep[see figure 7 in][]{2019MNRAS.486.5535M}.

This value of $q$ is close to the critical mass ratio $q_{\rm cr}$ at which a binary system may still possess positive superhumps. The exact value of $q_{\rm cr}$ is not known but is commonly accepted to be within 0.30--0.35, based on various theoretical and numerical studies  \citep[e.g.][]{2006MNRAS.371..235P,2011ApJ...741..105W,2015ApJ...803...55T}. 
There is a natural condition for the existence of $q_{\rm cr}$, namely the outer disk radius must be at least as large as the 3:1 resonance radius. The outer disk boundary is defined by the tidal interaction with the secondary star. At some distance from the primary, the tidal and viscous stresses become comparable and truncate the disk \citep{1977ApJ...216..822P,PapaloizouPringle77,IchikawaOsaki94}.
The outer disk boundary can be calculated as the largest streamline (the truncated orbit) in the restricted three-body problem that does not cross any other periodic orbit \citep{1977ApJ...216..822P}. We note that periodic orbits at large distances from the primary deviate from a circular Keplerian flow and cannot be characterised by a single parameter\footnote{To describe the largest streamlines, \citet{1977ApJ...216..822P} used two radial extrema $r_{\rm 1}$ and $r_{\rm 2}$, and the largest radius $r_{\rm max}$.} as these streamlines are elongated perpendicular to the line of centres of the two stars.

Adopting the approach by \citet{1977ApJ...216..822P}, \citet{2020AcA....70..313S} recently presented an approximation formula for the orbit-averaged radius of the truncated orbit:
\bea \label{eq:rtd}
       r_{\rm tid}/ r_{\rm Roche} & \approx & 0.830 + 0.860\,q - 4.974\,q^2 + 12.410\,q^3 \\ \nonumber
       && -14.842\,q^4 + 6.903\,q^5 .
\eea
Although the meaning of $r_{\rm Roche}$ is not given in the paper,
we can confirm, based on our calculations, that it is the volume radius of the Roche lobe of the white dwarf. It can be approximated using, e.g. a formula by \citet{1983ApJ...268..368E}
\be \label{eq:rl1}
       r_{\rm Roche} = r_{\rm L} = a\frac{0.49 q^{-2/3}}{0.6 q^{-2/3}+\ln(1+q^{-1/3})}.
\ee
Based on the fact that the intersection of the two relations (\ref{eq:epsr1}) and (\ref{eq:rtd}) gives $q$=0.22, \citet{2020AcA....70..313S} questioned the origin of positive superhumps due to the 3:1 resonance in higher $q$ systems. However, most of the recent SPH simulations, which treat tidal effects accurately, still show the appearance of the superhumps in binaries with a much higher mass-ratio, up to $q$$\approx$0.35 \citep[e.g.][]{2009MNRAS.398.2110W}
There might be several reasons for this disagreement. 

First, \citet{2020AcA....70..313S} in his analysis considered a simple case of circular orbits.  However, it is well known that the width of the 3:1 resonance depends strongly on the eccentricity 
\citep[e.g.][]{1993CeMDA..56..563H}, and all the orbits in the outer disk are highly non-circular. For example, using Eqn.~\ref{eq:rtd} for a system with $q$=0.300, one obtains the orbit-averaged circular radius $r_{\rm tid}/r_{\rm Roche}$ = 0.872. In reality, this orbit is highly elongated, which should significantly widen the resonance range. Indeed, using an approximation formula for $r_{\rm max}$ from \citet{2020A&A...642A.100N} 
    \begin{equation}
        {r_{\rm max} \over a} = 0.353+0.271\,e^{-3.045\,q}
    \end{equation}
one can obtain $r_{\rm max}/r_{\rm Roche}$ = 0.951, while $r_{\rm 1}/r_{\rm Roche}$ = 0.782 and $r_{\rm 2}/r_{\rm Roche}$ = 0.790.

Secondly, to calculate the truncated disk radius, \citet{1977ApJ...216..822P} adopted the inviscid three-body formalism. However, accretion disks are not collections of non-interacting test particles but are hydrodynamical flows where gas pressure effects could be potentially important. Indeed, \citet{Goodman93} demonstrated (see figure 3 in his paper) that if 
radial pressure gradients are not neglected then the disk in systems with $q$$\gtrsim$0.15 can expand beyond the Paczynski's radius. From the observational point of view, there exists observational evidence that accretion disks in some cataclysmic variables are extended all the way to $r_{\rm L}$ and that the matter can even escape from the disk in the orbital plane \citep[see, e.g.][]{2008PASJ...60L..23K,2017MNRAS.470.1960H,2020A&A...642A.100N,2020MNRAS.497.1475S,2024arXiv240511506Z}.

Thus, taking the above into account, in the following we assume that the positive superhumps observed in TT~Ari originate in a large precessing elliptical disk, which can be explained by the resonance model. Since the changes in the positive superhumps' period are much greater than the errors in its determination, the obtained interval of the mass ratio 0.29--0.37 cannot be considered as the result of statistical errors. The actual value of $q$, of course, does not change. Consequently, we suggest that the changes in the superhumps' period are associated with a change in the outer radius of the disk. It means that the lowest value of $\varepsilon^+$ corresponds to the case of the outer disk radius equal to the 3:1 resonance radius\footnote{More accurately, the disk has the radius at which the 3:1 resonance starts affecting the disk.}, and the increase of $\varepsilon^+$ is connected with the outer disk radius increasing. Here we point out again that the values 0.29 and 0.37 might be quite inaccurate as they were estimated in the extrapolation regime. 

Following \citet{2006MNRAS.371..235P} \citep[see also][]{1992ApJ...401..317L, 2001MNRAS.325..761M} we consider dependence of the positive superhump's period excess on the
tidal and pressure forces as 
\be \label{eq:epsr2}
\frac{\nu_{\rm pr}^+}{\nu_{\rm orb}}  =  \frac{\varepsilon^+}{1+\varepsilon^+}  = \frac{\nu_{\rm dyn}}{\nu_{\rm orb}} + \frac{\nu_{\rm press}}{\nu_{\rm orb}}, 
\ee
where $\nu_{\rm pr}^+$ is the frequency of the apsidal precession of the eccentric accretion disk.
\citet{1990PASJ...42..135H} derived an expression connecting the ratio of the precession frequency of an elliptical disk to the orbital frequency with the relative disk radius $r$ and the mass ratio $q$. It
can be represented as a series
\be \label{eq:dyn2}
\frac{\nu_{\rm dyn}}{\nu_{\rm orb}} =
    \frac{3}{4}\frac{q}{\sqrt{1+q}}r^{3/ 2}
    \sum_{n=1}^\infty c_n\,r^{2(n-1)}.
\ee
For our purposes, it suffices to confine ourselves to the first five terms of the series. The contribution of the higher-order terms is
 negligible. The values of the corresponding coefficients $c_n$ are taken from \citet{2003MNRAS.346L..21P} 
 and presented in Table\,\ref{tab:cn}.

According to \citet{1992ApJ...401..317L}, the precession frequency caused by the pressure effects can be presented as the following equation:
\be
        2\pi \nu_{\rm press} = - \frac{k^2 a_{\rm s}^2}{4\pi \nu_{\rm p}},
\ee
where $a_{\rm s}$ is the sound speed at the considered disk radius $R$,  $k$ is a radial wave number, and $2\pi \nu_{\rm p}$ is an angular rotation frequency of the disk matter at this radius. The radial wave number can be determined by using a pitch angle
$\beta$ of a spiral density waves generated by both tidal and pressure forces 
\be
          \tan \beta = \frac{1}{kR}.
\ee
Assuming Kepler orbital motion in the disk, we can  relate the frequency of the orbital motion to the orbital frequency:
\be
      \nu_{\rm p} = r^{-3/2}(1+q)^{-1/2} \nu_{\rm orb}.
\ee
Finally, the expression for ratio $\nu_{\rm press}/\nu_{\rm orb}$ can be obtained in the commonly accepted form
\be \label{eq:press2}
\frac{\nu_{\rm press}}{\nu_{\rm orb}} =
    -\frac{1}{2}\,r^{-1/2}(1+q)^{1/2}\left(\frac{1}{2\pi\nu_{\rm orb}}\times \frac{a_{\rm s}}{ a \tan 
    \beta} \right)^2.
    \ee  

\begin{table}
\caption {Coefficients $c_n$
\label{tab:cn}
}
\begin{center}
\large
\begin{tabular}{ccccc}
\hline
$c_1$ & $c_2$ & $c_3$ & $c_4$ & $c_5$ \\
\hline
\hline
            \noalign{\smallskip}
1 & $\frac{15}{8}$ & $\frac{175}{64}$ & $\frac{3675}{1024}$ & $\frac{72765}{16384}$ \\
            \noalign{\smallskip}
\hline
\end{tabular}
\end{center}
\end{table}

First of all, we estimate the mass ratio in the system assuming that the smallest value of $\varepsilon^+=0.078$ corresponds to
situation when the outer accretion disk radius equals to the resonance radius. 
From the simultaneous solution of the equations (\ref{eq:epsr1}) and (\ref{eq:epsr2}) we obtain that $q\approx 0.235$, if \textit{we 
ignore pressure effects}. The pressure effects decelerate the precession frequency, therefore, we expect that the mass ratio
has to be larger if we take into account pressure effects. 

The value of $\nu_{\rm press}/\nu_{\rm orb}$ depends not only on the disk radius and the mass ratio, but on the large
semi-axis of the system orbit $a$, the sound speed on the given radius $a_{\rm s}$, and the pitch angle of the spiral waves $\beta$.
\citet{2001MNRAS.325..761M} showed from her numerical simulations that the pitch angle is between 13 to 21 degrees. We use
this limits on $\beta$ in our further estimations. We also assume that the accretion disk in TT\,Ari is in a quasi-permanent hot state
during the observations, therefore, hydrogen is (almost) fully ionized and the plasma temperature on the outer disk radius is about
10$^4$\,K.  Therefore, we considered two values of the sound speed, 10 and 7.5 km\,s$^{-1}$. The first value approximately corresponds to the isothermal sound speed for  plasma temperature  10$^4$\,K, and we also consider the possibility that the plasma
temperature could be lower.  For the computation of the $a$ value, we assume that the white dwarf mass is 0.8 solar masses.  

Using the described above assumptions, we find that the $q$ value is between 0.24 ($a_{\rm s}$= 7.5 km\,s$^{-1}$ and $\beta=21\degr$) and 0.29  ($a_{\rm s}$= 10 km\,s$^{-1}$ and $\beta=13\degr$) if taking into account the pressure effects. In all the considered cases, the resonance radius $r_{\rm res}$ (see Eq.\ref{eq:epsr1}) is larger than the tidal radius $r_{\rm td}$ (see Eq.\ref{eq:rtd}) on 1-3\%.
We also investigated which outer disk radius is necessary to reach the maximum observed value of $\varepsilon^+=0.107$.
In all the considered cases, this radius is about 99\% of $r_{\rm L}$, determined by Eq.\,\ref{eq:rl1}.

Thus, we conclude that at $q=0.24-0.29$ the tidal resonance model as the cause of the precession of an elliptical disk is in agreement with the observations if we admit that the outer disk radius can be larger than the critical radius 
  $r_{\rm td}$, determined by  \citet{1977ApJ...216..822P}. In according to the model the increase in 
the period of positive superhumps is caused by increasing the outer radius of the disk. We also conclude that the upper possible value of $q\approx 0.29$ is in according to the estimation using Patterson's empirical relation
(\ref{eq:epsq}).

Based on the above calculations, we can estimate the masses of the components in the system.
There exist in the literature several empirical and theoretical relationships between the orbital period and the donor mass in cataclysmic variables, most of which predict a similar value of $M_2$ for the orbital period of TT~Ari, $P_{\rm orb}$=3.3 hours.
For example, the empirical relation by \citet{2005PASP..117.1204P} 
\be
     M_2 \approx 0.026\,P_{\rm orb}^{1.78}
     \,M_\odot,
\ee
where $P_{\rm orb}$ is expressed in hours, gives $M_2 \approx 0.22 M_\odot$. 
A more recent, semi-empirical broken-power-law donor sequence for cataclysmic variables from \citet{2011ApJS..194...28K} predicts $M_2$ to be $\approx$0.21 $M_\odot$. 
For further estimations, we assume that the mass of the donor star is  0.18$-$0.24 $M_\odot$, which gives the mass of the white dwarf to be 0.62$-$1\,$M_\odot$ at $q=0.24-0.29$.

\subsection{Tilted disk precession}

We showed above that the increase of the positive superhumps' period can be connected with the increase of the outer radius of the accretion disk. The positive superhumps dominate when their period is relatively short, 0.148--0.149 days \citep[see table 3 in ][]{2019MNRAS.489.2961B}. 
However, when the positive superhumps' period reaches 0.150--0.151 days, the domination of the negative superhumps begins. In particular, in the presented above observations, performed in 2021 and 2022, the negative superhumps dominate, although the relatively weak positive superhumps were also present, and their period was above 0.151~d. Therefore, in the framework of the interpretation of the negative superhumps as a precession of the tilted disk and the above suggestion that the increase of the positive superhumps' period is connected with the outer disk radius expansion, it leads to the suggestion that the disk becomes inclined (at least near the outer radius) after reaching some boundary radius.

We also note that the transition\footnote{Sometimes the negative and positive superhumps are observed simultaneously, although usually one of them dominates over the other. Therefore, there has to be some time interval when the dominance of one type of superhump modulations is replaced by the other type. It seems natural to call this time interval a transition period, and the corresponding event a transition.} from the positive to negative superhumps in TT~Ari has been associated with a brightness increase, which can also be related to an increase in the outer disk radius \citep[see, for example][]{2004AstLS}. 
The disk precession model of TT Ari presented above shows that the positive superhumps' period of about 0.150-0.151~d corresponds to the relative disk radius of about $r\approx 0.47-0.48$. We will call this relative outer disk radius a transition radius.

\begin{figure}
\includegraphics[width=0.99\columnwidth]{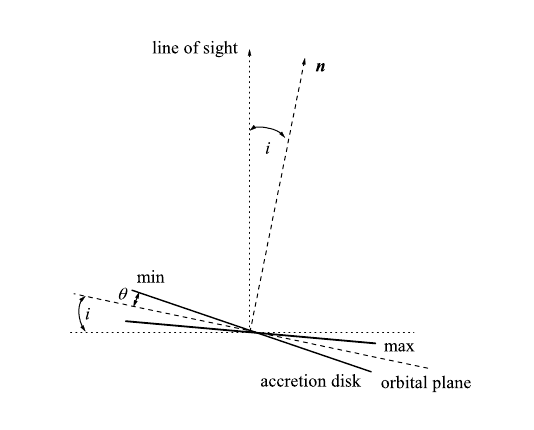}
\caption{Geometry of the tilted disk precession.
\textbf{\it n} is a normal to the orbital plane. The normal to the tilted accretion disk plane rotates
around {\bf$n$} with the negative superhumps period.} 
  \label{fig11}
\end{figure}

Brightness variability of the system when a tilted disk exists is most probably connected with changes in the apparent area of the disk. The apparent area of the tilted disk varies because of changing the inclination angle to the line of sight due to orbital motion and precession. In Fig.\,\ref{fig11} two positions of the accretion disk are shown. The positions correspond to two almost opposite orbital phases. In the first position  a brightness is maximal,
because the angle between the line of sight and the normal to the disk plane is minimal, $i-\theta$. In the second position, which shifted approximately on the orbital phase angle $\pi$, the brightness is minimal at the maximum value of the defined above angle, $i+\theta$. Here $\theta$ is the angle of inclination of the disk to the orbital plane of the system. The difference in the orbital phases of these two accretion disk positions is slightly less than $\pi$ due to nodal precession of the disk.

The light curves of the system show sine-like variations with a period close to the orbital one (see e.g. Fig.~\ref{fig4a}). This means that the angle $\theta$ is always less than $i$, otherwise the light curve would have been of a double-wave form.
Since the angle $i$ is small, the inclination of the disk to the orbital plane must also be small. We note, that the accretion disk can be inclined at the outer
radii only, with the disk plane coinciding with the orbital plane at the central parts.

\begin{figure}
\includegraphics[width=0.99\columnwidth]{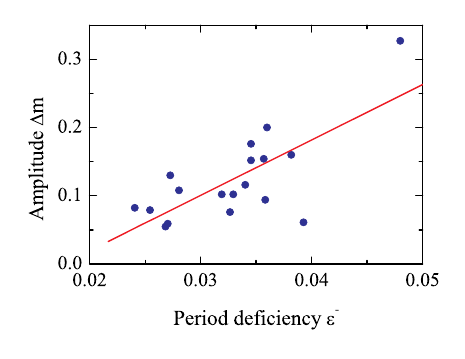}
\caption{Dependence of the amplitude of the system's brightness change on the negative superhumps period deficit $\varepsilon^-$ according to table 3 in \citet{2019MNRAS.489.2961B}.}
  \label{fig12}
\end{figure}

If we assume that the disk inclines to the orbital plane when the outer radius of the disk reaches the transition radius, it is logical to suggest that with a further increase in the disk radius, the inclination angle will also increase. The amplitude of brightness variability should also increase accordingly.
Indeed, the measurements given in table 3 of \citet{2019MNRAS.489.2961B} demonstrate that the amplitude of brightness variations of TT~Ari increases with the value of $\varepsilon^-$ (see Fig.\,\ref{fig12}).
Moreover, from the observations showing both types of superhumps simultaneously we see that with an increase of $\varepsilon^+$  between the TESS-21 and Kazan observations, the value of the corresponding negative superhumps period deficit $\varepsilon^- = (P_{\rm orb}-P_{\rm sh}^{-})/P_{\rm orb}$ also increased (see Table \ref{tab:per}). Thus, these observational facts support the idea that both an increase in the tilted angle of the disk and the increase in $\varepsilon^-$ are associated with an increase in the outer radius of the disk. 

The dynamics of a tilted disk have been studied in many papers.
As a rule, it is researched based on the precession of a point mass on an inclined orbit (with respect to the orbital plane) around a
white dwarf under the gravitation influence of the secondary component
\citep[see, for example,][and references therein]{2009ApJ...705..603M}.
By basing on \citeauthor{2009ApJ...705..603M}'s paper in which she considered the accretion disk as a collection of such particles uniformly precessing together, like a solid body, we obtained
a formula relating the ratio of the precession frequency to the orbital
frequency $\nu_{\rm pr}^-/\nu_{\rm orb}$ with disk radius and disk
inclination to the orbital plane $\theta$
\be \label{eq:epsneg}
    \frac{\nu_{\rm pr}^-}{\nu_{\rm orb}} = 
    \frac{\varepsilon^-}{1-\varepsilon^-} =
    \frac{15}{32}\frac{q}{\sqrt{1+q}}\,r^{ 3/2}\,\cos\theta \approx \varepsilon^-.
\ee

As mentioned above, the transition to a tilted disk possibly occurs at $r \approx 0.47-0.48$, and the relative disk radius cannot be larger than $r_{\rm L} \approx 0.49-0.504$ (for the largest and the smallest possible  $q$, 0.29 and 0.24, correspondingly). Therefore, according to (\ref{eq:epsneg}), the value of the frequency ratio turns out to be within $(0.034- 0.04)\,\cos \theta$. The period of negative superhumps varies from 0.1323 days \citep{2013AstL...39..111B, 2009A&A...496..765K} to 0.13424 days \citep{1999A&AK}, which results in frequency ratio values from 0.04 to 0.025. Therefore, the observed values are close to the computed values, because the angle $\theta$ is small and $\cos \theta \approx 1$.

The formal application of the expression (\ref{eq:epsneg}) to the observation data requires the value $\cos \theta > 1$ at the smallest observed values $P_{\rm ph}^-$ and $q< 0.29$.
 This may be due to the fact that the numerical coefficient in the equation (\ref{eq:epsneg})
is not known precisely, since it depends on the disk surface density distribution along the radius \citep[see, for example][]{1998MNRAS.299L..32L}. The value 15/32 was obtained in the homogeneous disk approximation, which is generally not true. However, we also suggest that the mass ratio is really closer to the found upper limit $q \approx 0.29$, or could be even larger if the pressure effects are more significant than we assumed. 

We note that the above consideration is applicable for the usual state of TT\,Ari with dominating of the negative superhumps like in 2021 and 2022.  The light curve observed in 2023 cannot be explained by the simple toy model presented above. One can speculate that the accretion disk was so highly tilted that the gas stream was meeting the disk at a radius much smaller than the outer disk radius, creating a bright spot inside the disk. This bright spot, which is visible during only half of the inclined disk precession period, can potentially generate the variability of the system brightness on a time scale of a few days.   

\subsection{Low frequency periodicities}

Here we speculate about possible reasons of the observed low frequency periodicities. The most interesting is the periodicity near the beat frequency $\nu^- - \nu^+$ observed in 2021 by the TESS observatory. 

This beat frequency can also be presented as a sum of the expected precession frequencies
$\nu_{\rm pr}^- + \nu_{\rm pr}^+$; the ratio $\nu_{\rm orb}/(\nu^- - \nu^+) \approx$\,1.107\,d/0.13755\,d $\approx 8$. Therefore, this beat frequency is close 
to be in a resonance with the orbital motion. We note that the sum of the nodal and the apsidal precession frequencies
of the orbit of the Moon is also in the resonance with the
orbital period of the Earth, 1/18.6\,yr + 1/8.8\,yr $\approx$ 1/6. This similarity can be an additional argument that the
observed negative and positive superhumps are connected with
the precession of the elliptical tilted accretion disk in according with the model suggested by \citet{1997PASP..109..468P}. On the other hand, the ratio
of the beat frequency with a low significance, observed in 2022 is close to be in the resonance 2:15, 1.033\,d/0.13755\,d $\approx$\,7.5.

In all the presented observations, there are two more or less stable periodicities at the periods 1.83-1.88\,d and 2.73-2.95\,d. It is possible to connect the first group of periods with the apsidal precession period at the resonance radius $r_{\rm res}$. Indeed, if we convert the period 1.88\,d to the positive superhumps period, we obtain the value, close to the minimal observed positive superhumps period in the system TT Ari, 0.1484\,d \citep[see table 3 in][]{2019MNRAS.489.2961B}, and the period excess $\varepsilon^+ \approx 0.078$. This value coincides with the value that we used in Sect.\,\ref{s321} for finding the resonance radius.
Therefore, we can suggest that the 3:1 resonance instability arises at the resonance radius, and we directly observe this precession period. The instability wave transfers to the outer disk radius with the decreasing precession period, and the observed negative superhumps period forms as a beat period between the precession period near the outer disk radius and the orbital period.
The direct reason for the flux variation can be the change in brightness of the place of the accretion stream and the accretion disk interaction.

The second group of the periods, 2.73-2.95\,d, can be connected with the nodal precession of the tilted accretion disk. These periods can be converted to the negative superhumps periods using relation (\ref{eq:eq1}), namely
0.1310-0.1314\,d. The corresponding period deficits,
$\varepsilon^-$, are 0.044-0.048. Such low negative superhumps periods were never observed and the expected
from (\ref{eq:epsneg}) values of $\varepsilon^-$ are smaller for the obtained $q$ range. Nevertheless,  
we can take courage and speculate that the considered group of the periods corresponds to the nodal precession of the tilted accretion disk with the radius close to the Roche lobe radius, whereas 
(\ref{eq:epsneg}) downplays the deficit of the period. It may to mean that the accretion disk always extends up to almost Roche lobe radius, but the
negative superhumps corresponds to the less radius, which can be connected with the place of the accretion stream and the accretion disk interaction.

\section{Summary and concluding remarks}
\label{sec:summary}

This paper presents a comparative analysis of photometric light curves of the cataclysmic variable TT Ari,  obtained by the TESS orbital observatory in 2021 and 2023, and derived from long ground-based observations in 2022, carried out with amateur telescopes.
The observational data and the performed analysis complement the recent extended studies of this cataclysmic variable star \citep{2019MNRAS.489.2961B,2022MNRAS.514.4718B}.

The system was in a bright state during all the observational runs. The character of variability was quite consistent throughout 2021-2022, with a predominance of a photometric period slightly shorter than the orbital one; this variability is interpreted as negative superhumps from a tilted accretion disk. However, the 2023 light curve exhibits much stronger modulations on a much longer time scale of a few days with an amplitude of up to 0.5 mag, compared to 0.2 mag in 2021. 
Previously dominant negative superhump modulations with a total amplitude of $\sim$0.10-0.12 mag were found at the beginning of the 2023 data set to have a much milder amplitude of $\sim$0.04 mag. Moreover, for the first time we observed that in the second half of the 2023 observational period, TT~Ari had experienced a transition to a state in which the amplitude of the negative superhumps  became even smaller.

In 2021-2022, in addition to the negative superhumps, TT~Ari also showed positive superhumps and modulations with the orbital period. The simultaneous presence of all three types of modulations has never previously been observed in TT~Ari. The detection of both positive and negative superhumps in the light curves strongly suggests that both the apsidal and the nodal precessions of the accretion disk in TT~Ari were found simultaneously. Thus, in 2021-2022, the disk was elliptical and also inclined to the orbital plane,
if the interpretation suggested by \citet{1997PASP..109..468P} is correct.

We attempted to detect low-frequency photometric modulations directly related to the nodal and/or apsidal precessions of the disk. No strong evidence for such modulations was found, although a periodicity close to expected for the nodal disk precession is seen in the first half of the 2023 observations. On the other hand, calculated periodograms show a few prominent low-frequency peaks, whose frequencies are close in different time intervals. The nature of these modulations is unclear.

The broad-band power spectra of the TESS and ground-based data are consistent with each other. In general, they can be described by red noise, a power-law with an exponent of $-2$, and with an additional signal superimposed on it, which makes a spectrum flat up to the frequency $\nu_c$= 90 d$^{-1}$ (72 d$^{-1}$ in 2023) and then falls off as a power-law at higher frequencies with an exponent of $-2.2$.
The difference in the level of this additional noise component and red noise is maximal near $\nu_c$, corresponding to a period of 16-20 minutes. On some nights, this resulted in the peaks in the periodograms at frequencies corresponding to 15$-$30 minutes.
However, the periodicity corresponding to these peaks appears to be short-lived. This confirms the conclusions made earlier by other authors \citep[see, e.g.][]{2014AcAS} that the quasi-periodic oscillations on time intervals of 10-60 minutes in TT Ari are short-lived and corresponding variability looks rather like a stochastic noise.

The obtained results were combined with those summarized by \citet{2019MNRAS.489.2961B} and compared with the predictions of theoretical models of accretion disk precession in binary systems.  We assumed that the resonance model suggested by \citet{1985A&A...144..369O} is acceptable for the description of the photometric properties of TT\,Ari, although it requires the resonant radius to be larger than the tidal radius. Thus, we assumed that the accretion disk in one or another form exists at radii greater than the boundary radius, as determined by \citet{1977ApJ...216..822P}. Application of the model of the elliptical accretion disk precession allowed us to find the mass ratio of the components in the system, $q$ is in the range 0.24-0.29. The uncertainty arises from the ambiguity of the pressure effects taken into account. The larger the pressure effect, the larger $q$.

The increase in the positive superhumps' period was interpreted as the outer disk radius expansion. It is important that the outer disk radius is less than the equivalent Roche lobe radius, even for the largest observed positive superhumps' period. It means that our assumption does not lead to any contradictions. However, if the superhumps' period significantly longer than the maximum period observed yet is found, it will mean that our assumption is wrong.  

The model of tilted disk precession predicts values of the precession period that are broadly consistent with observations.
The previously proposed suggestion that the amplitude and the period of negative superhumps also increase with an increase 
in the disk radius is also in agreement with the observations.

The use of various empirical relations between the mass of the secondary $M_2$ and the orbital period
allowed us to estimate the secondary mass $M_2$ to be in the range $0.18-0.24 \,M_\odot$, whereas the white dwarf mass $M_1$ to be  $0.62-1 M_\odot$.

\section*{Acknowledgments}
The authors thank the Referee for the very useful remarks and suggestions, and
Alexander Tokranov for his help with the English language. 
VFS thank the Deutsche Forschungsgemeinschaft (DFG) for financial support (grant WE 1312/59-1). 
VN acknowledges the financial support from the visitor and mobility program of the Finnish Centre for Astronomy with ESO (FINCA), funded by the Academy of Finland grant nr 306531.



\bibliographystyle{aa}
\bibliography{ttari_vn} 
\end{document}